\documentclass[]{aa}
\usepackage{graphicx}
\usepackage{txfonts}

\usepackage{natbib}

\begin{document}

\title{Acoustic oscillations in the field-free, gravitationally stratified cavities under
solar bipolar magnetic canopies}

\author{D. Kuridze\inst{1,2} \and T.V. Zaqarashvili\inst{2} \and B. M. Shergelashvili\inst{1,2,3} \and S. Poedts\inst{1}}

\institute{Center for Plasma Astrophysics, K.U.Leuven, 200 B,
B-3001, Leuven, Belgium,\\ \email{dato.k@genao.org} \and Georgian
National Astrophysical Observatory at the Faculty of Physics and
Mathematics, I. Chavchavadze State University, Al. Kazbegi ave. 2a,
0160 Tbilisi, Georgia, \and Institute for Theoretical Physics, K.U.
Leuven, Celestijnenlaan 200 D, B-3001, Leuven, Belgium, 
}

\date{Received / Accepted}

\abstract {}
{The main goal is to study the dynamics of the gravitationally
stratified, field-free cavities in the solar atmosphere, located
under small-scale, cylindrical magnetic canopies, in response to
explosive events in the lower-lying regions (due to granulation,
small-scale magnetic reconnection, etc.).}
{We derive the two-dimensional Klein-Gordon equation for isothermal density
perturbations in cylindrical coordinates. The equation is first solved by a standard
normal mode analysis in order to obtain the free oscillation spectrum of the cavity. Then,
the equation is solved in the case of impulsive forcing associated
to a pressure pulse specified in the lower-lying regions.}
{The normal mode analysis shows that the entire cylindrical cavity
of granular dimensions tends to oscillate with frequencies of 5-8
mHz and also with the atmospheric cut-off frequency. Furthermore,
the passage of a pressure pulse, excited in the convection zone,
sets up a wake in the cavity oscillating with the same cut-off
frequency. The wake oscillations can resonate with the free
oscillation modes, which leads to an enhanced observed oscillation
power.}
{The resonant oscillations of these cavities
explain the observed power halos near magnetic network cores and
active regions.}


\titlerunning{Acoustic oscillations under solar bipolar magnetic canopies}

\authorrunning{Kuridze et al.}

\maketitle

\section{Introduction}

Observations show a high spectral power of oscillations
($\nu>5\;$mHz) in the vicinity of active regions. The observed
velocity power maps are sometimes referred to as `photospheric power
halos' (Braun et al. 1992; Brown et al. 1992; Hindman \& Brown 1998;
Jain \& Haber 2002; Muglach et al. 2005; Moretti et al. 2007;
Nagashima et al. 2007; Hanasoge 2008). The power enhancement is
observed only in the Doppler velocity power maps and not in
continuum intensity (Hindman \& Brown 1998; Jain \& Haber 2002;
Muglach et al. 2005; Nagashima et al. 2007). The enhanced power
spectra peak at the period of $\sim3\;$min.

In addition, observations show that the chromospheric active regions and
magnetic network elements are surrounded by ``magnetic shadows'',
which lack oscillatory power in the higher frequency range (McIntosh
and Judge 2001; Krijger et al. 2001; Vecchio et al. 2007). From
the observations it seems reasonable to conclude that both the photospheric
power halos and the chromospheric magnetic shadows reflect
the same physical process.

In our previous paper (Kuridze et al. 2008), we proposed an
explanation of the photospheric power halos as the effect of a
magnetic canopy on the wave dynamics. We showed that the field-free
cavity regions under the magnetic canopy can trap high-frequency
acoustic oscillations, leading to the observed increased
high-frequency power in the photosphere, while the lower-frequency
oscillations are channeled upwards in the form of
magneto-acoustic waves (Erd{\'e}lyi et al. 2007; Srivastava et al.
2008). However, those calculations have been performed without
taking into account the gravitational stratification, which is important at the
photospheric level.

In the present paper, we study acoustic oscillations in the
stratified, cylindrical field-free cavity regions under the magnetic
canopy as a possible explanation of the power halo phenomenon. There
are various analytical and numerical investigations of the acoustic
wave propagation in isothermal stratified atmospheres (e.g., Lamb
1908; Rae \& Roberts 1982; Fleck \& Schmitz 1991; Kalkofen et al.
1993; Sutmann et al. 1998; Roberts 2004). However, almost all these
calculations involve simple 1D models (for vertically propagating
acoustic waves), because 2D and 3D models necessarily include
internal gravity waves, which considerably complicates the analysis.
In order to avoid further complications due to the gravity waves, it
is possible, however, to only consider {\em isothermal propagation},
which automatically neglects gravity waves. This enables us to study
the 2D propagation of only isothermal acoustic waves or pulses in a
stratified atmosphere. Here, we study normal isothermal acoustic
oscillations and the propagation of acoustic pulses in a cylindrical
field-free cavity by solving the 2D Klein-Gordon equation in
cylindrical geometry. Unlike the 1D case, the two-dimensional
solution allows us to investigate the variation of the
density/velocity perturbation amplitude in the azimuthal direction.
First, we use a standard normal mode analysis to study the spectrum
of possible acoustic oscillations in the cavity region. Then, we
solve the Klein-Gordon equation with impulsive forcing in the
photospheric region.

The outline of the paper is as follows. In Sect.~2 we present the
basic hydrodynamic equations for the problem of the propagation of
acoustic waves in a stratified cavity medium. Section~3 describes
the normal mode approach  and the resulting oscillation spectrum in
the cavity. In Sections 5 and 6 we analyze the response of the
cavity region to a pressure pulse. The results obtained are
discussed in Sect.~7.

\section{The wave equation}

We use the ideal hydrodynamic equations for a gravitationally stratified
field-free cavity:

\begin{equation}
{\partial \rho \over \partial t}+\nabla \cdot (\rho {\bf v})=0,
\end{equation}

\begin{equation}
\rho\left({\partial {\bf v} \over \partial t}+{\bf v}\cdot\nabla
{\bf v} \right) =-\nabla p+\rho {\bf g},
\end{equation}

\begin{equation}
{\partial p\over\partial t}+{\bf v} \cdot\nabla p+{\gamma p\nabla
\cdot {\bf v}} =0,
\end{equation}
\bigskip
where ${\bf v}$ is the fluid velocity, $p$ and $\rho$ denote the
pressure and density, respectively, $\gamma$ is the ratio of specific
heats, and ${\bf g}$ is the gravitational acceleration.

The considered acoustic cavity under the cylindrical magnetic canopy
has a semicircular form (Fig.1). Therefore, it is convenient to
consider the equations (1-3) in a cylindrical coordinate system. For
simplicity, we study the 2D case, i.e.\ perturbations polarized in
the $(r,\phi)$ plane, where $r$ is the radial coordinate and $\phi$
is the azimuthal angle.

In this cylindrical (or polar) coordinate system, the equilibrium
pressure and density are related as follows:

\begin{equation}
{\partial p_{01}\over \partial r}=g_r\rho_{01},
\end{equation}

\begin{equation}
{1 \over r}{\partial p_{01}\over \partial \phi}=g_{\phi}\rho_{01},
\end{equation}
where $p_{01}$,  $\rho_{01}$ denote the equilibrium pressure and
density and

$$
g_r=-g \sin\phi,
$$
$$
g_{\phi}=-g \cos\phi,
$$
are the $r-$ and $\phi-$components of the gravitational acceleration vector,
${\bf g}$, respectively.

The Eqs.~(4-5) then easily lead to the equilibrium values of the pressure
and the density in the cavity:

\begin{equation}
p_{01}=\tilde{p}_{01}e^{-r\sin \varphi /\Lambda},
\end{equation}
\begin{equation}
\rho _{01}=\tilde{\rho }_{01}e^{-r\sin \varphi /\Lambda},
\end{equation}
where
\begin{equation}
\Lambda={p_{01}\over g\rho _{01}},
\end{equation}
is the pressure scale hight. The atmosphere is considered to be
isothermal. Therefore, this pressure scale height is constant.


In the considered cylindrical (polar) coordinate system, the linearized hydrodynamic equations for adiabatic fluctuations can be written as:

\begin{equation}
\
\frac{\partial \rho _{1}}{\partial t}-\rho _{01}\frac{\sin \varphi }{\Lambda}u_{r}%
-\rho _{01}\frac{\cos \varphi }{\Lambda}u_{\varphi }+\frac{\rho _{01}}{r}\frac{\partial }{\partial r}%
\left( ru_{r}\right) +\frac{\rho _{01}}{r}\frac{\partial u_{\varphi
}}{\partial \varphi }=0 ,\
\end{equation}

\begin{equation}
\
\rho _{01}\frac{\partial u_{r}}{\partial t}=-\frac{\partial p_{1}}{\partial r%
}-g\sin \varphi\rho _{1}, \
\end{equation}

\begin{equation}
\ \ \rho _{01} \frac{\partial u_{\varphi }}{\partial
t}=-\frac{1}{r}\frac{\partial p_{1}}{\partial \varphi }-g\cos
\varphi\rho _{1},
\end{equation}

\begin{equation}
\ \frac{\partial p_{1}}{\partial t}-c_{0}^{2}\frac{\partial \rho
_{1}}{\partial t}+\rho _{01}\frac{c_{0}^{2}}{g}(\sin \varphi u_{r}
+\cos \varphi u_{\varphi } )\omega _{b}^{2}=0, \
\end{equation}
where $c^2_0={\gamma p_{01}/\rho_{01}}$ is adiabatic uniform sound
speed and $\omega_b$ denotes the Brunt-V\"{a}is\"{a}l\"{a}
frequency:
\begin{equation}
\omega^2_b={(\gamma-1)g\over \gamma \Lambda}.
\end{equation}

These equations describe the acoustic and internal gravity waves
propagating in a stratified atmosphere. In principle,
the cavity under a magnetic canopy may trap both types of waves.
Unfortunately, the simultaneous consideration of both wave types is
very complicated, especially from an analytical point of view.
However, by adopting specific approximations, we can consider each of
the wave types separately. The limit of incompressibility, e.g., neglects the acoustic
branch and only gravity waves remain in this case. On the other hand, the limit
of isothermal propagation neglects the gravity branch of the spectrum and in this
case only the acoustic branch remains. The latter limit means that the temperature
exchange is so rapid that any temperature fluctuation in the
perturbations is zero. This is achieved when $\gamma$ approaches
unity.

We use the approximation/limit of isothermal propagation in the remaining part of this paper.  For
isothermal perturbations, Eqs.~(9-12) lead to the following wave equation:
$$
c_{0}^{2}\frac{\partial ^{2}\rho _{1}}{\partial r^{2}}+\left( g\sin
\varphi
+\frac{c_{0}^{2}}{r}\right) \frac{\partial \rho _{1}}{\partial r}+\frac{%
c_{0}^{2}}{r^{2}}\frac{\partial ^{2}\rho _{1}}{\partial \phi ^{2}}
$$

\begin{equation}
+\frac{g}{r%
}\cos \varphi \frac{\partial \rho _{1}}{\partial \varphi }-
\frac{\partial ^{2}\rho _{1}}{\partial t^{2}}=0.
\end{equation}

Equation~(14) can be written in a more convenient form by using the following
transformation:
\begin{equation}
{\rho }_{1}=\tilde{\rho} _{1}e^{-\lambda r\sin \varphi },
\end{equation}
where the parameter $\lambda$ can be chosen in such a way that the
resulting wave equation has the form of a Klein-Gordon equation in
the cylindrical/polar coordinate system. This procedure is analogous
to the well-known transformation in a 1D Cartesian coordinate system
(Kahn 1990; Sutmann et al. 1998).

By substituting Eq.~(15) into Eq.~(14) we obtain:
$$
c_{0}^{2}\frac{\partial ^{2}\tilde{\rho }_{1}}{\partial
r^{2}}+\frac{c_{0}^{2}}{r}\frac{\partial \tilde{\rho }_{1}}{\partial
r}+\frac{c_{0}^{2}}{r^{2}}\frac{\partial ^{2}\tilde{\rho
}_{1}}{\partial \varphi ^{2}}-\frac{\partial ^{2}\tilde{\rho
}_{1}}{\partial t^{2}}-\left(2 \lambda c_{0}^{2}-g\right)\sin
\varphi\frac{\partial \tilde{\rho }_{1}}{%
\partial r}
$$
\begin{equation} -\left(2 \lambda c_{0}^{2}-g\right){\cos
\varphi \over r}\frac{\partial \tilde{\rho }_{1}}{%
\partial \phi}+(\lambda^{2}c_{0}^{2}-\lambda g) \tilde{\rho }%
_{1}(r)=0,
\end{equation}
from which we find the following condition on $\lambda$
\begin{equation}
2\lambda c_{0}^{2}-g=0,
\end{equation}
or
\begin{equation}
\lambda={g \over 2c_{0}^{2}}={1\over 2\Lambda}.
\end{equation}
Using this condition, we obtain
\begin{equation}
\frac{\partial ^{2}\tilde{\rho }_{1}}{\partial r^{2}}+{1\over
r}\frac{%
\partial \tilde{\rho }_{1}}{\partial r}+{1\over r^2}
\frac{\partial ^{2}\tilde{\rho }_{1}}{\partial \phi^{2}}-{1\over
c^2_{0}} \left [\frac{\partial ^{2}\tilde{\rho }_{1}}{\partial
t^{2}}+\omega^2_{ac}\tilde{\rho_1}\right ]=0,
\end{equation}
where  $\omega^2_{ac}$ refers to the acoustic cut-off frequency
\begin{equation}
\omega^2_{ac}={g\over 4\Lambda}={c_0^2\over 4\Lambda^2}.
\end{equation}

Equation~(19) is the Klein-Gordon equation in cylindrical/polar
coordinates. This equation governs the isothermal propagation of
acoustic waves (and pulses) in the considered gravitationally
stratified medium. The very dynamic convective layer located under
the cavity can excite quasi-harmonic wave trains as well as acoustic
pulses. Clearly, the cavity will respond in different ways to these
different types of perturbations. We study two different responses
of the cavity to such different perturbations from below. First, we
consider the cavity as resonator for acoustic waves and, using the
standard normal mode approach, we study the spectrum of possible
wave harmonics trapped in the canopy. Next, we consider the
propagation of a pressure pulse, which can be caused in the lower
regions e.g.\ due to the eruption of new granular cells or by some
small-scale magnetic reconnection processes.

\section{Normal mode analysis}

\begin{figure}[]
\vspace*{1mm}
\begin{center}
\includegraphics[width=7.0cm]{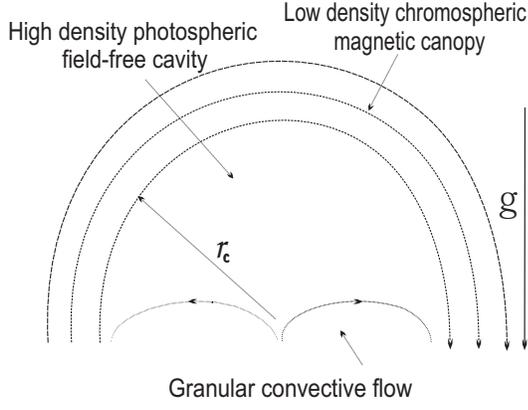}
\end{center}
\caption{Simple schematic picture of small-scale magnetic canopy
overlying a field-free cavity with gravitational stratification,
\vec{g}.}
\end{figure}

A standard Fourier analysis of Eq.~(19) with respect to time, leads to the
equation:
\begin{equation}
r^{2}\frac{\partial ^{2}\tilde{\rho }_{1}}{\partial r^{2}}+r\frac{%
\partial \tilde{\rho }_{1}}{\partial r}+ \frac{\partial ^{2}\tilde{\rho
}_{1}}{\partial \phi^{2}}+ r^{2}\left( \frac{\omega
^{2}}{c_{0}^{2}}-{\omega^2_{ac}\over c^2_0}\right)\tilde{\rho }_{1}
=0.
\end{equation}

$$\frac{\omega ^{2}}{c_{0}^{2}}-\frac{g}{4\Lambda
c_{0}^{2}}\equiv k_1^2,$$ which gives
$$\omega^2=k_1^2c_0^2+\omega^2_{ac}.$$

\begin{figure}[]
\vspace*{1mm}
\begin{center}
\includegraphics[width=9.0cm]{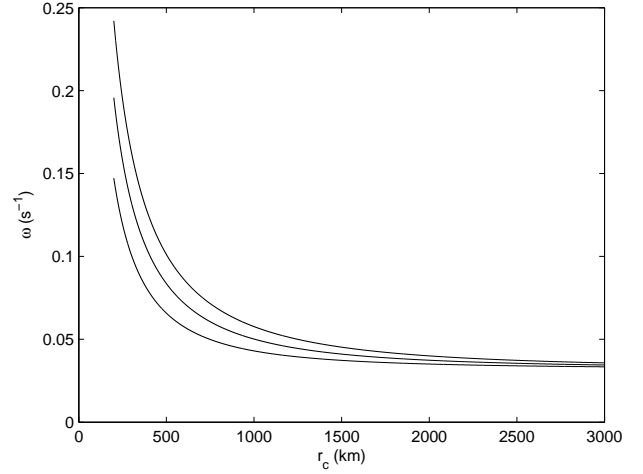}
\end{center}
\caption{The oscillation frequency, $\omega$, versus the cavity
radius, $r_c$, for the first zeros of the $m=1$ (bottom line), $m=2$
(middle line) and $m=3$ (top line) harmonics.}
\end{figure}

We consider the solution of Eq.~(21) in three different cases
depending on the size of the circular frequency $\omega$ with
respect to  $\omega_{ac}$.

When $\omega=\omega_{ac}$, equation (21) leads to the Laplace
equation in cylindrical coordinates. The solution of this equation
inside the canopy (i.e.\ inside the semicircle $r=r_c$, where $r_c$
is the cavity/canopy boundary (see Fig.~1)) with the condition
$\tilde\rho_1=\tilde\rho_{c}(\phi)$ (i.e.\ a Dirichlet condition)
along the boundary is given by (Morse \& Feshbach 1953)
\begin{equation}
{\tilde\rho}_1=\sum_{m=0}^{\infty}[A_m\cos(m\phi)+B_m\sin(m\phi)]\left({r\over
r_c}\right)^m,
\end{equation}
where
\begin{equation}
A_m={2\over\pi}\int_{0}^{\pi}\tilde\rho_{c}(\alpha)\cos(m\alpha)d\alpha,
\end{equation}
and
\begin{equation}
B_m={2\over\pi}\int_{0}^{\pi}\tilde\rho_{c}(\alpha)\sin(m\alpha)d\alpha.
\end{equation}
This means that one of the free oscillation modes in the cavity
oscillates at the cut-off frequency for any $m$ (which plays the
role of azimuthal wave number). For the photospheric sound speed of
$7.5\;$km/s, the acoustic cut-off frequency is of the order of
$0.031\;$s$^{-1}$ with a corresponding period of $\sim 200\;$s.

In the case $\omega>\omega_{ac}$, Eq.~(21), again after a Fourier
analysis with respect to the azimuthal direction and with azimuthal
wave number $m$, transforms to a Bessel equation with the general
solution
\begin{equation}
{\tilde\rho}_1=c_{1}J_m(k_1r)+c_{2}Y_m(k_1r),
\end{equation}
where $J_m$ and $Y_m$ are the bessel functions of the first and
second kind, respectively, and $c_1$ and $c_2$ are arbitrary constants.

When $\omega<\omega_{ac}$, then a similar Fourier analysis
transforms Eq.~(21) to the modified Bessel equation, with the
general solution
\begin{equation}
{\tilde\rho}_1=c_{3}I_m(k_1r)+c_4K_m(k_1r),
\end{equation}
where $I_m$ and $K_m$ are the modified Bessel functions of the first
and second kind, respectively, and $c_3$ and $c_4$ are arbitrary
constants. The density perturbations must be finite at $r=0$, which
yields the condition $c_2=c_4=0$. We also need a second condition at
the upper boundary of the cavity in order to find a unique solution
of equation (21). The correct approach is to find an analytical
solution of the magnetohydrodynamic equations in the overlying
magnetic canopy region and then merge it with the solution (25) at
the canopy/cavity interface ($r=r_c$). This would yield the exact
oscillation spectrum of the cavity, which may reveal some wave
leakage. Unfortunately, we could not get an analytical solution of
the complete cylindrical MHD equations for the gravitationally
stratified plasma in the canopy region. However, we can exploit the
fact that the density quickly decreases from the photosphere to the
chromosphere. According to the VAL-C atmospheric model (Vernazza et
al. 1981), there is the sharp density gradient in the low area of
the chromosphere (Fontenla et al. 1990). Therefore, it is justified
to approximate the sharp gradient between the photospheric and the
chromospheric densities as a discontinuity at the canopy/cavity
interface.
The density is then much higher in the cavity than in the overlying
magnetic canopy (Fig.1). This allows the use of a free boundary
condition, i.e.\ $\tilde \rho_1=0$, at the cavity/canopy interface,
which yields an approximate spectrum of acoustic oscillations in the
cavity. The influence of overlying canopy can be studied by complete
numerical simulations, but this is not the scope of the present
paper and should be done in the future. The free boundary condition
leads to the expressions:

\begin{equation}
c_1 J_m(k_1r_c)=0,  \\ for \\ \omega > \omega_{ac},
\end{equation}
\begin{equation}
c_3 I_m(k_1r_c)=0, \\ for \\ \omega < \omega_{ac},
\end{equation}
The first condition gives oscillatory solutions (oscillating with
respect to the $r-$coordinate), while the second condition is
satisfied only when $c_3=0$.  Therefore, Eq.~(28) yields the trivial
solution of Eq.~(21), and in the following we will concentrate on
the oscillatory solution connected with the condition (27) only.

The spectrum of acoustic oscillations in the cavity can then be
deduced from the zeros of the solution $J_m(z)$, which (as it is known from the
theory of Bessel functions) are all real when $m\geq -1$, and they can be
easily found in tables (see e.g.\ Abramowitz \& Stegun 1967).

Figure~2 shows the dependence of the frequencies of the acoustic
oscillations on the cavity radius, $r_c$, for the first zero of
$J_m(z)$ for the $m=1,2,3$ harmonics. It is evident from this figure
that the frequency decreases with increasing $r_c$, as can be
expected from physical considerations. This means that these
cavities, with typical granular radii $r_c=400-800\;$km, support
oscillations with frequencies in the range $0.07-0.05\;$s$^{-1}$
(i.e. with periods of $1.5-2\;$min). For considerably large cavity
radii, the frequencies of all the harmonics tend to the cut-off
value.


It seems that cylindrical cavities support higher frequency
oscillations than typical quiet Sun regions. This is because in
cylindrical cavities, the oscillations are partly along the
$\phi-$direction due to the small $k_1$ term for the first zero of
Bessel function in Eq.~(27), which prevents their leakage upwards
and leads to their trapping in the cavities. On the other hand, the
regions without overlying canopies (e.g., nonmagnetic quiet Sun
regions outside the network cores) can not trap the high frequency
oscillations, because they may propagate upwards. Therefore, it is
expected that the high frequency oscillations are trapped around the
magnetic network cores and active regions where the cylindrical
magnetic canopies can be formed. The periods of the lower order
harmonics in cavities with a radius of $1000\;$km (Fig.~2),
correspond to the observed spectrum of acoustic oscillations
($5-8\;$mHz) (Moretti et al. 2007) (see their Fig.~1).

The free boundary condition, which is used to obtain the oscillation
spectrum, does not lead to wave leakage in the canopy region.
However, we may explore the problem using the wave propagation along
a narrow sector around the vertical ($\phi\approx\pi/2$) direction.

Let us first consider the unperturbed current-free cylindrical
magnetic field in the canopy region expressed as (see Kuridze et al.
2006) $B_{\phi}=B_{\phi0}(r_c/ r)$, where $B_{\phi0}$ is the
magnetic field strength at canopy/cavity interface. For simplicity,
we consider the zero-$\beta$ approximation ($\beta=8\pi
p_{02}/B^2_{\phi0}\approx 0$), where $p_{02}$ is the plasma pressure
in the canopy. For this configuration, the solution of
magnetohydrodynamic equations for the radial velocity is found in
form of the Hankel function of the first kind (Kuridze et al. 2008):


\begin{equation}
{\hat u_{r2}}=ic_3rH_{m/2}(k_2r),
\end{equation}
where $c_3$ is an arbitrary constant and
$k_2={{\omega}/{(2v_{A})}}$. Here, $v_{A}= {B_{\phi0}/\sqrt{4\pi
\rho_{02}}}$ is the Alfv{\'e}n speed and $\rho_{02}$ is the plasma
mass density in the canopy area. The radial velocity component in
the narrow sector around the vertical direction of the cavity can be
obtained combining Eqs.~(10), (15) and (25) as
\begin{equation}
{\hat u_{r1}}=e^{r/2\lambda}{ic_{1}\over {\tilde\rho_{01}
\omega}}\left (c^2_{0}J^{\prime}_m(k_1r)+{g\over 2}J_m(k_1r)\right
).
\end{equation}
The continuity of the velocity and the total pressure
perturbations at the canopy/cavity interface then leads to the following
dispersion equation:
\begin{equation}
e^{r/\Lambda}{{\omega^2}\over v^2_{A}}{\rho_{01}\over
\rho_{02}}{{J_m(k_1r_c)}\over
J_m^{\prime}(k_1r_c)+J_m(k_1r_c)/2\Lambda}= -{{
H_{m/2}^{\prime}(k_2r_c)}\over H_{m/2}(k_2r_c)}.
\end{equation}

The dispersion relation (31) is a transcendental equation for the
complex frequency $\omega$. A non-vanishing imaginary part of
$\omega$ indicates wave leakage from the field-free cavity into the
overlying magnetic canopy. The question then arises how important
this wave leakage is. The analytical solution of Eq.~(31) is
complicated. Therefore, we apply numerical techniques to solve it.

\begin{figure}[t]
\vspace*{1mm}
\begin{center}
\includegraphics[width=10.0cm]{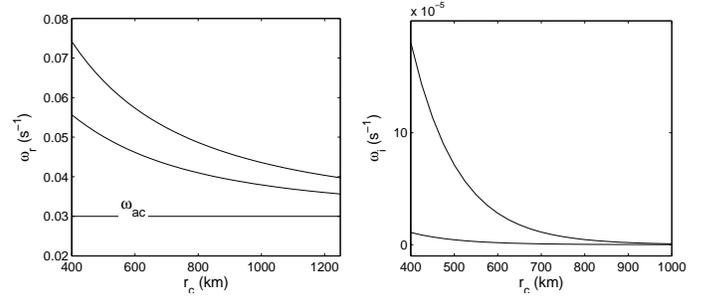}
\end{center}
\caption{Real $\omega_r$ ($m=1$ bottom line, $m=2$ top line on the
left panel) and imaginary $\omega_i$ ($m=1$ top line, $m=2$ bottom
line on the right panel) parts of wave frequency vs the radius of
field-free cavity $r_c$. The straight line on the left panel is the
solution with the cut-off frequency.}
\end{figure}

The numerical solution of Eq.~(31) shows that the dispersion
equation has a real solution with a cut-off frequency
$\omega=\omega_{ac}$, and complex solutions with very small
imaginary parts. Figure~3 shows the dependence of the real (left
panel) and the imaginary (right panel) parts of the frequency on the
cavity size $r_c$. The real part of the frequency decreases with
increasing $r_c$, as expected. The damping time of the oscillations
$t_d=1/\omega_i \sim 10^3-10^5 \, T_0$ ($T_0=2\pi/\omega_r$) is very
large, which indicates there is almost no wave leakage in the canopy
region. This means that the first few harmonics are trapped in the
cavity. It must be mentioned that the frequencies of first few
harmonics (left panel of the Fig.~3) are almost identical to the
frequencies obtained with the free-boundary condition (Fig.~2).
Thus, the oscillation spectrum does not depend significantly on the
choice of the boundary conditions.

It is interesting to find out what happens if the cylindrical canopy
is replaced by a horizontal magnetic field (this could be the case
in the quiet Sun regions far from the network cores). We have
analyzed the spectrum of the field free photospheric area under a
horizontal magnetic canopy. In this configuration there are no
trapped high frequency ($\omega>\omega_{ac}$) oscillations along the
vertical direction. The damping times of the vertically propagating
waves are about $t_d\sim2-7\;T_0$, which indicates there leaky
nature. On the other hand, horizontally propagating waves (with
large horizontal wave numbers) have real frequencies. But since
their velocities are almost horizontal they can not be seen in
Doppler velocity power maps. Any initial pulse or wave train will
quickly be dispersed in the horizontal direction. Therefore, these
modes can not form the observed high-frequency halos.
%

\section{Excitation by a pressure pulse}

The solar photosphere is very dynamic and it contains many different
types of impulsive sources. For example, the eruption of new
granules, magnetic field reconnection events just under the solar
surface, and various other explosive events may take place there.
Therefore, the excitation of pressure and velocity pulses seems to
be quiet common under photospheric conditions. In this section, we
study the propagation of such a pressure pulse (which corresponds to
a density pulse in the isothermal approximation) in the cavity
region. For simplicity, we assume that the pulse has a
$\delta$-function shape in space and time.


Mathematically this corresponds either to the solution of Eq.~(19) with
impulsive initial and boundary conditions or to the solution of that
equation with an additional term modeling the impulsive external forcing. The
equation can then be written as (with the driving force term in the right-hand side)
$$
\frac{\partial ^{2}\tilde{\rho }_{1}}{\partial r^{2}}+{1\over
r}\frac{%
\partial \tilde{\rho }_{1}}{\partial r}+{1\over r^2}
\frac{\partial ^{2}\tilde{\rho }_{1}}{\partial \phi^{2}}-{1\over
c^2_{0}} \left [\frac{\partial ^{2}\tilde{\rho }_{1}}{\partial
t^{2}}+\omega^2_{ac}\tilde{\rho_1}\right ]
$$
\begin{equation}
=-{4\pi\delta(t-t_0)\delta(r-r_0)\delta(\phi-\phi_0)\over r},
\end{equation}
where $r_0$, $\phi_0$  denotes position of the source (Fig.~4).
Equation~(32) governs the propagation of the pulse, which is set at
$r=r_0$, $\phi=\phi_0$ at the time $t=t_0$.

If we consider the case when
\begin{equation}
\tilde\rho_1(r,t,\phi|r_0,t_0,\phi_0)={{\partial
\tilde\rho_1(r,\phi,t|r_0,t_0,\phi_0)}\over
\partial r}=0
\end{equation}
for $t<t_0$, then $\tilde{\rho}_1$ is the Green's function for
Eq.~(32) and has a form (Morse \& Feshbach 1953)

$$
\tilde\rho_1(r,t,\phi|r_0,t_0,\phi_0)={A_0\delta(t-t_0-R/c_0)\over
R}
$$
\begin{equation}
-{A_0\omega_{ac}J_1\left[\omega_{ac}\sqrt{(t-t_0)^2-{(R/c_{0})^2}}\right]\over
c_0 \sqrt{(t-t_0)^2-(R/c_{0})^2}} H(t-t_0-R/c_0),
\end{equation}
where $A_0$ is a constant,
$$
R=\mid {\bf r}-{\bf r_0}
\mid=\sqrt{r^2+r^2_0-2rr_0\cos(\phi-\phi_0)},
$$
and $H$ denotes the Heaviside step function.

\begin{figure}[t]
\vspace*{1mm}
\begin{center}
\includegraphics[width=8.0cm]{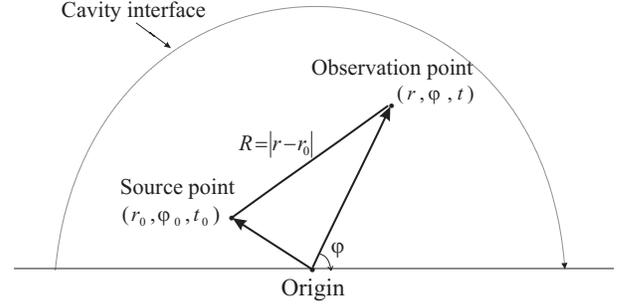}
\end{center}
\caption{Source point, observation point at $t>t_0$ and
canopy/cavity interface.}
\end{figure}

\begin{figure}[]
\vspace*{1mm}
\begin{center}
\includegraphics[width=9.3cm]{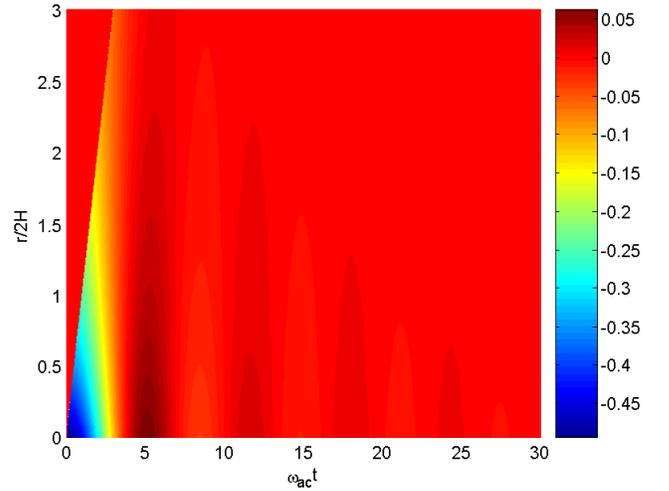}
\end{center}
\caption{The density fluctuation after the propagation of pulse as a
function of time and radial coordinate at $\phi=\pi/4$. 
The time $t$ is normalized by cut-off frequency and $r$ is
normalized by the photospheric scale height.}
\end{figure}

The first term in expression (34) describes the propagation of the
pulse. It is followed by a wake which is given by the second term in
Eq.~(34). This is a demonstration of a well-known result (Rae \&
Roberts 1982; Fleck \& Schmitz 1991; Kalkofen et al. 1993; Sutmann
et al. 1998; Roberts 2004; Zaqarashvili \& Skhirtladze 2008), viz.\
the presence of a wavefront which moves away from the point ($r_0,
\phi_0$) with the speed $c_0$. The disturbance ahead of the
wavefront is at rest, but behind the wavefront the medium begins to
oscillate at the cut-off frequency $\omega_{ac}$.

Using the expressions (15), (18) and (34) the density perturbation
for the wake oscillations takes the form  (here $t_0=0$ is chosen
without loss of generality):

\begin{equation}
\rho _{1}= -{A_0\omega_{ac}e^{-r\sin \varphi /2\Lambda}\over c_0
\sqrt{t^2-(R/c_{0})^2}}J_1\left[\omega_{ac}\sqrt{t^2-{(R/
c_{0}})^2}\right]H(t-R/c_0).
\end{equation}

An asymptotic form of the density perturbation can be obtained using the
expression of the Bessel function $J_1$ that holds for large
arguments $(t>>R/c_0)$ (Abramowitz \& Stegun 1967). We then obtain
\begin{equation}
\rho _{1}=-{\sqrt{2\over \pi}}{A_0\sqrt{\omega_{ac}}\over
c_0}{e^{-r\sin \varphi /2\Lambda}\over
{t^{3/2}}}\cos(\omega_{ac}t-3\pi/4).
\end{equation}
After the passage of the pressure pulse, the medium density thus
oscillates at the cut-off frequency and this oscillation decays
asymptotically as $1/t^{3/2}$.  In fact, the pressure pulse will
excite the wide spectrum of oscillations, but the higher frequencies
harmonics escape upwards and the lower frequency harmonics are
evanescent thus only the oscillations at the cut-off frequency
remain.


It is interesting to study what happens with the velocity
perturbations in this asymptotic case ($t>>R/c_0$). Combining
Eqs.~(10), (11), and (34) we obtain

\begin{equation}
{\partial u_r\over \partial t}={A_0c_0\sqrt{\omega_{ac}}\over
2\Lambda\tilde \rho_0}\sqrt{2\over
\pi}{\cos(\omega_{ac}t-3\pi/4)\over t^{3/2}}e^{r\sin \varphi
/2\Lambda}\sin\phi,
\end{equation}
and
\begin{equation}
{\partial u_{\phi}\over \partial t}={A_0c_0\sqrt{\omega_{ac}}\over
2\Lambda\tilde \rho_0}\sqrt{2\over
\pi}{\cos(\omega_{ac}t-3\pi/4)\over t^{3/2}}e^{r\sin \varphi
/2\Lambda}\cos\phi.
\end{equation}

The integration of Eqs.~(37) and (38) with respect to time and using
the conditions $u_r=0$ and $u_{\phi}=0$ for $t\rightarrow\infty$, we
obtain (see the Appendix for the detailed calculations):
\begin{equation}
\hat u_{r}=-\sqrt{{2\over\pi}}{A_0\sqrt{\omega_{ac}}\over \tilde
\rho_0 \Lambda}{e^{r\sin \varphi /2\Lambda}\sin\phi\over
{t^{1/2}}}\cos(\omega_{ac}t-3\pi/4),
\end{equation}
and
\begin{equation}
\hat u_{\phi}=-\sqrt{{2\over\pi}}{A_0\sqrt{\omega_{ac}}\over \tilde
\rho_0 \Lambda}{e^{r\sin \varphi /2\Lambda}\cos\phi\over
{t^{1/2}}}\cos(\omega_{ac}t-3\pi/4),
\end{equation}
where $\hat u_r=u_r/c_0$ and $\hat u_{\phi}=u_{\phi}/c_0$.
Therefore, the propagation of the pressure pulse also sets up a
velocity wake oscillating with the cut-off frequency. Note, that
this velocity oscillation decays with time as $1/t^{1/2}$, i.e.\
much more slowly than the decay of the density wake. Thus the
oscillations may easily persist during a few tens of cut-off periods
(i.e.\ during $\sim30-60\;$min), which is much longer than the
expected time interval between consecutive pulses (which can be
estimated as the granular life-time, i.e.\ $\sim10\;$min).
Therefore, these wakes of consecutive pulses may gradually
strengthen the oscillations and thus may lead to the observed power
halos.


Figure~5 shows a plot of the density fluctuation, given by Eq.~(35)
(the time is normalized to the cut-off frequency, $t'=t\omega_{ac}$,
and the radial coordinate is normalized by the scale height,
$r'=r/2\Lambda$). It is seen that the propagation of the pulse is
followed by a weak amplitude wake which oscillates with the cut-off
period. The amplitude of both the pulse and the wake decrease with
height/time. However, it must be mentioned that the normalized
perturbation of the density, $\rho_1/\rho_0$, increases with $r$.
The normalized amplitude of the density perturbation in the
asymptotic form, can be written as
\begin{equation}
\hat\rho_1=-{\sqrt{2\over \pi}}{A_0\sqrt{\omega_{ac}}\over
\tilde\rho_0c_0}{e^{r\sin \varphi /2\Lambda}\over
{t^{3/2}}}\cos(\omega_{ac}t-3\pi/4),
\end{equation}
where
$$
\hat\rho_1={\rho_1\over \rho_0}.
$$
The normalized amplitude of the asymptotic density perturbation
depends on the direction of propagation (i.e.\ on the
$\phi-$coordinate). The wake has its maximum amplitude in the
vertical direction (at $\phi=\pi/2$) and its amplitude decreases for
inclined directions (see Fig.~6).


Figure~7 shows the dependence of the asymptotic velocity wake on the
azimuthal angle, $\phi$. It is clear that the radial component has a
maximum amplitude in the vertical direction, while the azimuthal
velocity component has a maximal amplitude close to $45^{0}$.

\begin{figure}[t]
\vspace*{1mm}
\begin{center}
\includegraphics[width=9cm]{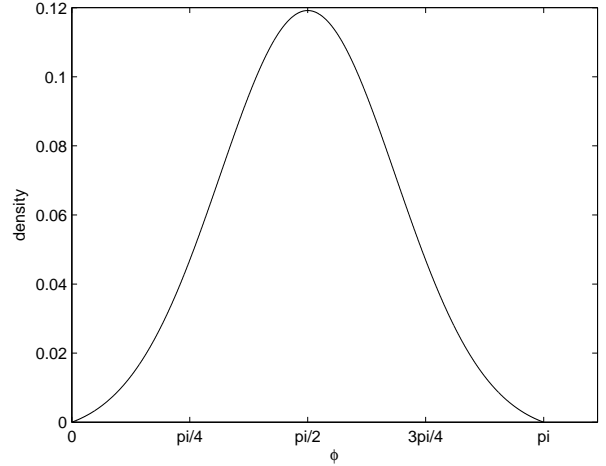}
\end{center}
\caption{Normalized asymptotic amplitude of the density wake versus
the azimuthal angle $\phi$ at $r'=2$, $t'=100$ (the time $t$ is
normalized to the cut-off period and $r$ is normalized to the scale
height).}
\end{figure}

\section{The energy propagation}

Our calculations show clearly that at $t=0$ the energy is concentrated in the initial
pressure pulse, and that it propagates away as time elapses. The
initial pulse, which travels at the sound speed, carries most of the
energy injected into the medium. We can write down an expression for the asymptotic total
energy density for the wake oscillation, which is the sum
of the kinetic and acoustic potential energy densities
$$
E\thicksim \hat{u}^2_r+\hat{u}^2_{\phi}+\hat\rho_1^2=K^2_1{e^{r\sin
\varphi /\Lambda}\over {t}}\cos^2(\omega_{ac}t-3\pi/4)
$$
\begin{equation}
+K^2_2{e^{r\sin \varphi /\Lambda}\over
{t^{3}}}\cos^2(\omega_{ac}t-3\pi/4)
\end{equation}
where,
$$K_1=\sqrt{{2\over\pi}}{A_0\sqrt{\omega_{ac}}\over \tilde \rho_0
\Lambda},$$
and
$$K_2={\sqrt{2\over \pi}}{A_0\sqrt{\omega_{ac}}\over \tilde\rho_0 c_0}.$$
It is clear from the expression (42) that the potential part of the
wake energy decays much faster than the kinetic part. Therefore,
after some time only the velocity oscillations remain, while the
density oscillations have decayed. Furthermore, the total energy is
the largest in the vertical direction (at $\phi=\pi/2$) and the
smallest in the horizontal direction (at $\phi=0$). It means that
the energy of the wake oscillations is concentrated near the
vertical direction (see Fig.~8). Therefore, the wake, which remains
after the passage of the pressure pulse, tends to oscillate along
the vertical direction. The same result was found by Bodo et al.
(1999). The total energy of the wake decays in time as $\thicksim
1/t$.

\section{Discussion and conclusions}

\begin{figure}[]
\vspace*{1mm}
\begin{center}
\includegraphics[width=9.3cm]{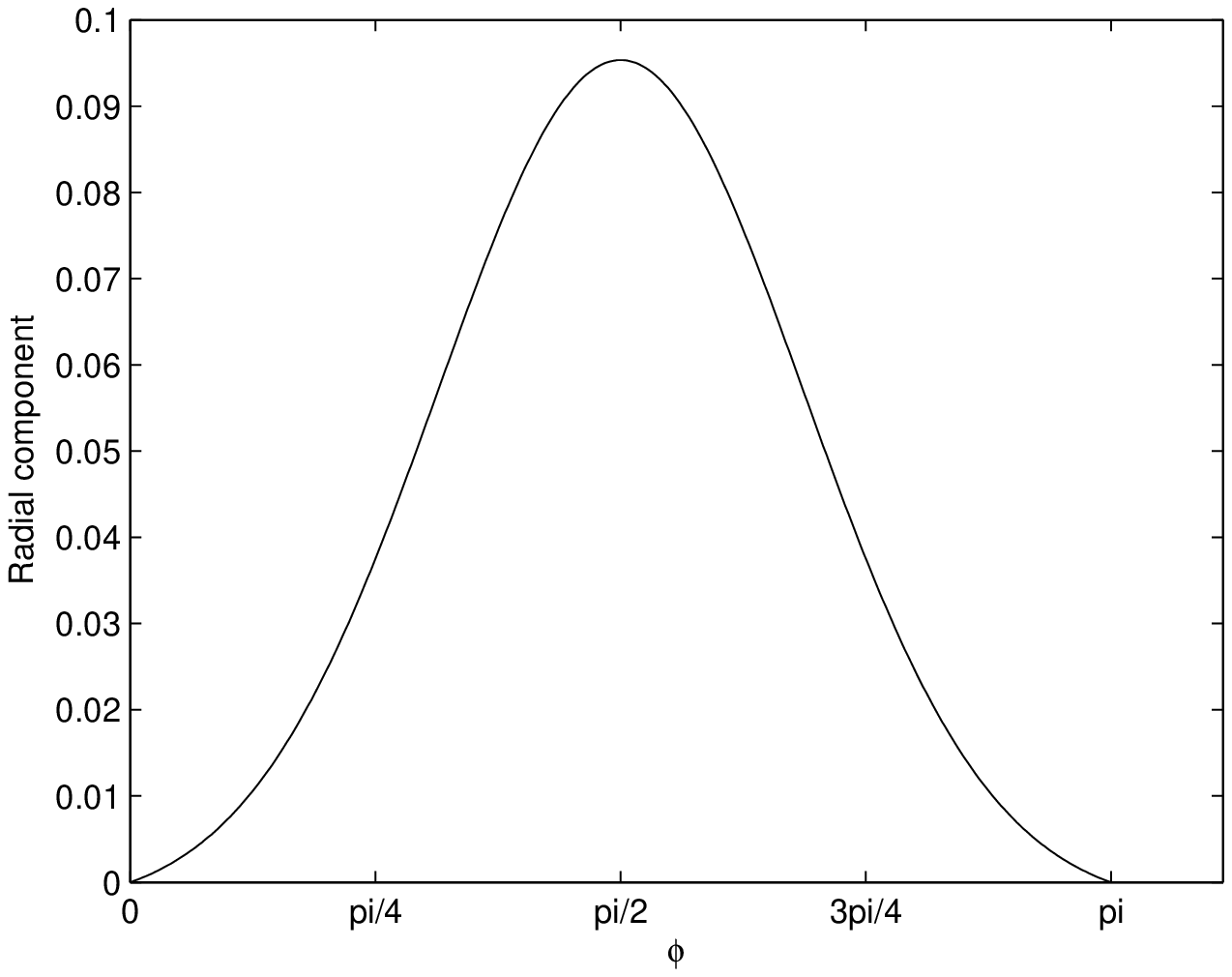}
\includegraphics[width=9.3cm]{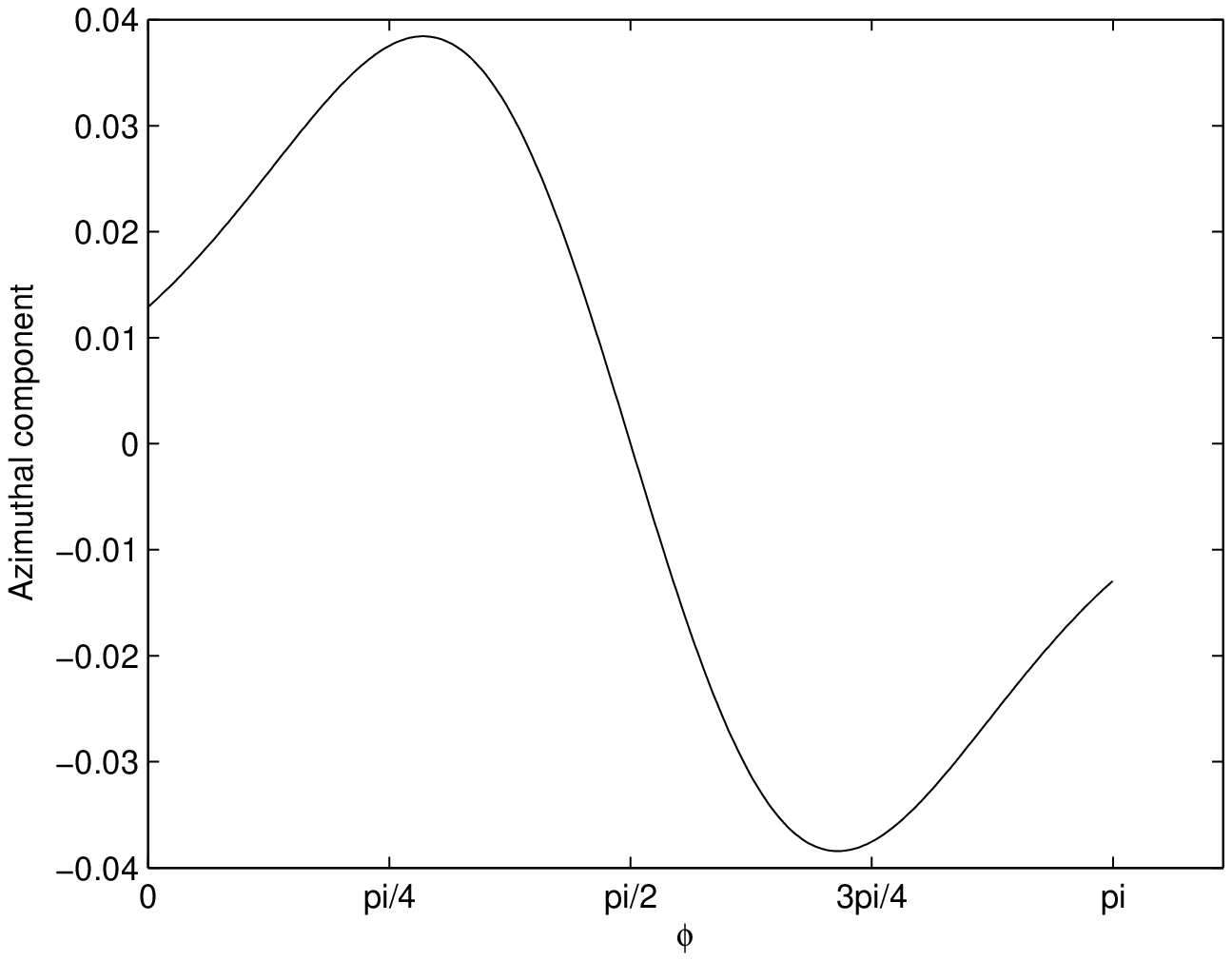}
\end{center}
\caption{Normalized asymptotic amplitudes of the velocity
perturbation components versus $\phi$ at $r'=2$, $t'=100$ (the time
$t$ is normalized to the cut-off period and $r$ is normalized to the
scale height).}
\end{figure}

\begin{figure}[]
\vspace*{1mm}
\begin{center}
\includegraphics[width=9.3cm]{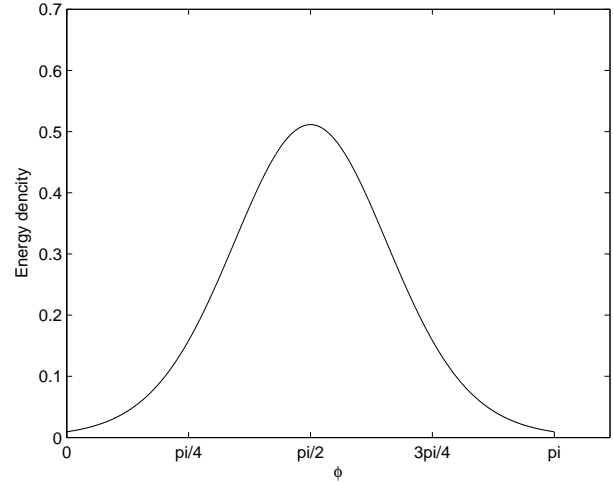}
\end{center}
\caption{Normalized total energy density with respect to $\phi$. It
is clearly seen that the energy is concentrated near the vertical
direction. }
\end{figure}

We have carried out a linear perturbation analysis of the
two-dimensional wave propagation in a stratified, field-free cavity,
which is located under a chromospheric cylindrical magnetic canopy.
We consider the higher density in the cavity than in the overlying
magnetic canopy region (due to the sharp density gradient between
the photosphere and the chromosphere). Then the free boundary
condition at canopy/cavity interface leads to the spectrum of
acoustic oscillation in the cavity. We also have examined the
response of the cavity to the propagation of a pressure pulse, which
can be excited in the lower regions due to various sources,
including e.g. small-scale magnetic reconnection events. It has been
shown that the cavity, which has a radius comparable to the granular
dimensions ($\sim$ 500 km), supports trapped oscillations with
periods of $1.5-2\;$min (cf.\ Fig.~2,~3). However, one of the
oscillation modes in the cavity is the oscillation with the cut-off
frequency. Furthermore, the frequency of the acoustic modes, also
has the tendency to approach the cut-off value for large scale
granular size (cf.\ Fig.~2). On the other hand, it has been shown
that a pressure pulse propagating through the cavity sets up a wake,
oscillating at the cut-off frequency. The wake oscillation will
resonate with a free oscillating mode of the cavity, which may lead
to an enhanced acoustic power at the same frequency.

There have been several different explanations of power halos
proposed in the literature: (i) the enhancement of acoustic emission
by some unknown source (Braun et al. 1992; Brown et al. 1992; Jain
\& Haber 2002), (ii) incompressible oscillations, such as Alfv\'en
waves or transverse kink waves, in magnetic tubes (Hindman \& Brown
1998),  (iii) the interaction of acoustic waves with the overlying
magnetic canopy (Muglach et al. 2005; Kuridze et al. 2008) and (iv)
a change of the spatial-temporal spectrum of the turbulent
convection in the magnetic field (Jacoutot et al. 2008). However,
observations show a lack of power halos in the intensity maps
(Hindman \& Brown 1998; Jain \& Haber 2002; Muglach et al. 2005;
 Nagashima et al. 2007), which needs an adequate explanation.
These observations challenge the model of the power halos as due to
acoustic waves. However, our model easily explains the discrepancy.

We suggest that the surroundings of the magnetic network cores and
the active regions consist of many small-scale closed magnetic
canopy structures (McIntosh \& Judge 2001; Schrijver \& Title 2003).
Granular motions transport the magnetic field at the boundaries and
consequently create field-free cylindrical cavity areas under the
canopy (Fig.~1). The pressure pulses excited in lower region
propagate through the cavity and leave behind wakes oscillating at
the cut-off frequency. These wake oscillations resonate with the
free oscillation modes of the cavity. Therefore, the cavities
accumulate acoustic oscillations with the observed frequency.
However, the amplitude of the density perturbation in the wakes
decreases in time faster (viz.\ as $t^{-3/2}$) than the amplitude of
velocity perturbations (viz.\ as $t^{-1}$). Therefore, the density
perturbations quickly decay and only the velocity oscillations
remain. This explains why the power haloes are seen only in the
velocity power maps.

Thus, our model suggests that the high-frequency acoustic
oscillations should be trapped around the magnetic network cores and
the active regions, where acoustic cavities can be formed under
small-scale cylindrical magnetic canopies. The deduced oscillation
periods are within $1.5-2\;$min for cavities with granular sizes,
but additionally the cavities should oscillate with the photospheric
cut-off frequency. However, the picture will be different far away
from the network cores and active regions, where the cylindrical
bipolar magnetic fields are probably replaced by almost a horizontal
magnetic canopy which has been used as a model for a long time
(Evans \& Roberts, 1990).  In this case, the higher frequency
oscillations propagate upwards (only trapped cut off modes appear)
and, consequently, power halos can not be formed. The oscillation at
cut-off frequency should have a small amplitude there, as the
rectangular vertical cavity does not have an oscillation mode with
this frequency. The wake oscillating at cut-off frequency should be
observed there, but with a smaller amplitude as almost the whole
initial energy is carried away by an initial pulse or a wave train.
Therefore, it is the cylindrical structure of the field-free regions
(see Fig.~1), which helps to trap acoustic waves as the oscillations
occur partly along the $\phi-$direction. This explains why the
''power halos'' and ``magnetic shadows'' are observed only near the
quiet-Sun chromospheric magnetic network cores (McIntosh and Judge,
2001; Krijger et al., 2001; Vecchio et al., 2007).


It must be noted that intensity halos are visible in the magnetic
canopy regions in some observations  (Moretti et al. 2007). However,
this is not inconsistent with our model. As mentioned before, our
calculation does not include the exact solution in the overlying
magnetic canopy region due to mathematical difficulties. On the
other hand, the free oscillations of the cavity/canopy interface may
excite oscillations in the overlying region as observed by Moretti
et al. (2007). Additionally, the oscillations of the cavity may
excite transverse waves in overlying magnetic canopy, which can be
checked by observations of the magnetic field oscillation in the
power halo regions.

\begin{acknowledgements}

This work has been supported by Georgian National Science Foundation
grant GNSF/ST06/4-098. D.K. is grateful for kind hospitality at the
Center for Plasma Astrophysics, K.U.Leuven during his visit when the
significant parts of the work has been developed.
These results were obtained in the framework of the projects
GOA/2009-009 (K.U.Leuven), G.0304.07 (FWO-Vlaanderen) and
C~90347 (ESA Prodex 9).

\end{acknowledgements}

\begin{appendix}

\section{Derivation of the velocity components}

Integration of Eqs.~(37) and (38) gives:

\begin{equation}
u_r =e^{r\sin \varphi /2\Lambda}\sin\phi{A_0c_0\omega_{ac}\over
2\Lambda\tilde \rho_0}\sqrt{2\over \pi}\int{\cos(t'-3\pi/4)\over
t'^{3/2}}dt',
\end{equation}

\begin{equation}
u_{\phi}=e^{r\sin \varphi /2\Lambda}\cos\phi{A_0c_0\omega_{ac}\over
2\Lambda\tilde \rho_0}\sqrt{2\over \pi}\int{\cos(t'-3\pi/4)\over
t'^{3/2}}dt',
\end{equation}
where $t'=\omega_{ac}t$.

Integral in  Eqs. (A.1), (A.2) can be solved as

$$
I =\int{\cos(t'-3\pi/4)\over t'^{3/2}}dt'=
$$
$$
=cos{3\pi\over 4}\int{{\cos t'\over t'^{3/2}}}dt'+sin{3\pi\over
4}\int{{\sin t'\over t'^{3/2}}}dt'
$$
$$
=\cos{3\pi\over 4}\left[ -{2\cos t'\over t'^{1/2}}-2\sqrt{2\pi}
S\left(\sqrt{2t'/\pi}\right)+F_1\right]
$$
$$
+\sin{3\pi\over 4}\left[ -{2\sin t'\over t'^{1/2}}+2\sqrt{2\pi}
C\left(\sqrt{2t'/\pi}\right)+F_2\right]
$$
$$
=-{2\over t'^{1/2}}\left[\cos{3\pi\over 4}\cos t'+\sin{3\pi\over
4}\sin t' \right]
$$
$$
+2\sqrt{\pi}\left[S\left(\sqrt{2t'/\pi}\right)+C\left(\sqrt{2t'/\pi}\right)\right]+{\sqrt{2}\over
2}\left(F_2-F_1\right)
$$
$$
=-2{\cos(t'-3\pi/4)\over
t'^{1/2}}+2\sqrt{\pi}\left[S\left(\sqrt{2t'/\pi}\right)+C\left(\sqrt{2t'/\pi}\right)
\right]
$$
\begin{equation}
+{\sqrt{2}\over 2}\left(F_2-F_1\right),
\end{equation}
where $F_1$, $F_2$ are the integration constants, and $S$, $C$ is
the Fresnel sine and cosine integral functions, respectively.

Fresnel functions tend to $1/2$ for large arguments, therefore we
have
\begin{equation}
I=-2{\cos(t'-3\pi/4)\over t'^{1/2}}+2\sqrt{\pi}+{\sqrt{2}\over
2}\left(F_2-F_1\right).
\end{equation}

$u_r, u_{\phi}$ should tend to zero when $t'\rightarrow\infty$,
therefore
\begin{equation}
{\sqrt{2}\over 2}\left(F_2-F_1\right)=-2\sqrt{\pi},
\end{equation}
which gives
\begin{equation}
I=-2{\cos(t'-3\pi/4)\over t'^{1/2}}.
\end{equation}

Substitution of the expression (A.6) into Eqs. (A.1), (A.2) gives
Eqs. (39) and (40).

\end{appendix}




\end{document}